\documentstyle[preprint,aps]{revtex}
\newcommand{\be}{\begin{equation}}
\newcommand{\ee}{\end{equation}}
\newcommand{\bea}{\begin{eqnarray}}
\newcommand{\eea}{\end{eqnarray}}

\newcommand{\p}{\partial}

\tightenlines 

\begin{document}

\draft
\preprint{\small   }

\title{Apparent horizons, black hole entropy and \\loop quantum gravity}

\author{Viqar Husain\footnote{
Email address: husain@physics.ubc.ca}}

\address{\baselineskip=1.4em Department of Physics and Astronomy,\\
University of British Columbia,\\
6224 Agricultural Road \\
Vancouver, BC V6T1Z1, Canada}

\maketitle 

\begin{abstract}

In recent work on black hole entropy in non-perturbative quantum
gravity, an action for the black hole sector of the phase space is
introduced and (partially) quantized. We give a number of observations
on this and related works. In particular we show that (i) the entropy
calculation applies without change to generally covariant theories
having no black hole solutions, (ii) the phase space constraint used
to select the black hole sector is not the apparent horizon equation,
which is the natural phase space constraint separating trapped and
untrapped regions on an initial data surface, and (iii) there appears
to be at least one other phase space constraint which leads to the
conclusion that the entropy associated with a bounding two-dimensional
surface is proportional to its area.

\end{abstract}
\bigskip
\pacs{PACS numbers:  }

Among the latest developments in the non-perturbative approach to
quantum gravity utilizing the triad-connection canonical variables
\cite{aa,js,rs,ai,al,tt}, is a set of ideas that help to answer the 
question: what are the microscopic degrees of freedom of a black-hole? 
Very briefly, the answer in this approach is that black hole microstates
are certain ``surface'' degrees of freedom residing on the horizon of
a black hole. Counting of these degrees of freedom yields an entropy
$S_{BH}$ proportional to the area $A$ of the black hole event horizon
for large black holes, in accord with the Beckenstein-Hawking formula;
the proportionality constant in the proposal contains a parameter,
which it is argued may be fixed to give $S_{BH} = A/4$ in Planck
units.

The initial concrete realization of the idea that surface degrees of
freedom residing on the event horizon may be identified with black
hole microstates was given by Carlip \cite{carlip} for the
three-dimensional black hole. \footnote{For a recent reanalysis of
this work see \cite{carlip2}.} In four dimensions, the stage for
calculating black hole entropy as a counting of surface degrees of
freedom was set by work of Smolin \cite{lee} on the Beckenstein bound,
and of Rovelli \cite{rovarea} and Krasnov \cite{krasarea} on counting
the degeneracy of spin network states associated with fixed area.

The four-dimensional black hole entropy calculation due to Ashtekar,
Baez, Corichi and Krasnov \cite{aetal} in the loop quantum gravity
approach is similar in spirit to these earlier works. It contains a
concrete proposal for an action for the ``black hole sector'' of
general relativity. This is used as the starting point for canonical
quantization.

In this note we give an outline review of the background and main
ingredients of this proposal, and then give a number of observations
on them. These are: (I) a nearly identical analysis may be
carried out in theories without ``dynamics,'' which have no black hole
solutions, with the resulting entropy having a different
interpretation, namely, that an entropy (proportional to area) may be
associated to any surface, (II) the phase space condition used in
selecting the ``black hole sector'' of the gravitational phase space
in \cite{aetal} is not the apparent horizon equation, which is the
phase space condition giving the boundary separating trapped and
untrapped regions on a spatial slice, and (III) other boundary
conditions may also give a surface entropy proportional to the surface
area.

The non-perturbative quantum gravity approach starts with a classical
first order ``self-dual'' action in which the gravitational degrees of
freedom are the SL(2,C) vierbein $e_a^{AA'}$, and self-dual part of
the SL(2,C) connection $A_{aB}^A$. This connection has only chiral
2-spinor components, and is therefore complex \cite{actions}. Boundary
terms may be added to the bulk action as required by the physical
context to guarantee a well-defined variational principle.

The main ingredients in the black hole entropy calculation are as
follows; (the reader is referred the original papers \cite{aetal}, and
references therein for further details).

(i) Attention is restricted to non-static and asymptotically
flat black hole spacetimes $M$ with an inner boundary (the event
horizon $H$). A specific phase space condition is specified on this
boundary, and is incorporated into the following proposed action for
black hole spacetimes:
\be 
S[e,A] = {1\over 8\pi G_N}\int_M {\rm Tr}\left[ e\wedge e\wedge F(A)\right]
+ {1\over 8\pi G_N} {A_S\over 4\pi} \int_H {\rm Tr}\left[ A\wedge dA 
+ {2\over 3} A\wedge A \wedge A\right],
\label{action}
\ee 
where $A_S$ is the area of the 2-surface boundary $S$ where spatial 
slices $\Sigma$ intersect the horizon. The boundary condition 
which makes this action functionally differentiable is 
\footnote{For details of the origin of this and other conditions 
see \cite{aetal}.} 
\be 
\underline{F}_{ab} = - {2\pi\over A_S}\ \underline{\sigma}_{ab},
\label{bc1}
\ee
where $\sigma_{ab}= (e\wedge e)_{ab}$ is the dual of the momentum
$E^a_i$ conjugate to $A_a^i$, and the underline denotes restrictions
of the fields to $S$; $i=1,2,3$ is the SU(2) index).  Since the
surface term in this action contains time derivatives, there is an
additional surface contribution to the phase space Poisson brackets,
and hence dynamical surface degrees of freedom.
 
(ii) This action leads to the complex phase space configuration
variable $A_a^i = \Gamma_a^i - iK_a^i$, where $\Gamma_a^i$ is the
spatial spin connection and $K_a^i$ is the extrinsic curvature. At
this stage a change to real variables is made \cite{barb,holst,imm}:
this variable is replaced with the real phase space variable $^\gamma
A_a := \Gamma_a^i - \gamma K_a^i$, where $\gamma$ is a real parameter
(the Barbero-Immirzi parameter \cite{imm}).  Its conjugate momentum
variable is $^\gamma \tilde{E}^a = (1/ \gamma)\tilde{E}^a$. The
horizon boundary condition on the two-surface $S$ in $\Sigma$ is now
\be
^\gamma\underline{F}_{ab} 
= - {2\pi\gamma\over A_S}\ ^\gamma\underline{\sigma}_{ab}^{AB},
\ee
where both sides of the equation are real. This ($\gamma$ dependent) 
equation is a part of the conditions used to identify the black hole 
sector of the phase space, and is the one imposed quantum mechanically. 
(The remaining conditions in \cite{aetal} are not directly relevant 
for the purposes of this paper.)
 
(iii) The kinematical constraints arising from the above action, which
generate the SU(2) Gauss law and spatial diffeomorphisms, can be
quantized; the kinematical quantum states $\psi[A]$ are spin network
states \cite{spin1,spin2}. These are basically the generalization of
Wilson lines to graphs, with edges labelled by SU(2) representations,
and group index contractions at the vertices via generalizations of
6-j and 9-j etc. symbols. In the black hole context, the full Hilbert
space is a direct product of ``volume'' and ``surface'' states:
$\psi[A]=\psi_V[A]\bigotimes\psi_S[A]$.

(iv) The horizon boundary condition in (2) is imposed as the 
following quantum condition to select out kinematical black hole states: 
\be 
\left( 1{\otimes} {2\pi\gamma\over A_S}\hat{
\underline{F}}_{ab}\cdot r
+ \hat{\underline{\sigma}}_{ab} \cdot r
{\otimes} 1\right)\, \Psi_V {\otimes} \Psi_S\, 
= 0 
\label{qbc1}
\ee
where $r$ is a fixed internal vector. The first and second terms act
entirely on surface and volume states respectively. The volume states
are spin network states, and the surface states are states of
Chern-Simons theory with sources provided by spin network edges
intersecting the surface \cite{aetal}.

(v) Spin network states are also eigenstates of the area operator
\cite{area1,area2}. Therefore, the relevant solutions of (\ref{qbc1})
are those spin network states which are compatible with the fixed area
$A_S$ in (\ref{qbc1}). The entropy is calculated by counting the
number of solutions of (\ref{qbc1}) with this added restriction. This
counting gives entropy proportional to $A_S$.

We now give three observations and related results on this black 
hole entropy calculation:

(I) These steps are entirely kinematical in the sense that the
Hamiltonian constraint, or its consequences, do not enter any of the
calculations: the space of states used are solutions to the Gauss and
spatial diffeomorphism constraints, and the phase space condition
(\ref{bc1}) is also independent of the bulk dynamics (as any phase
space condition other than the Hamiltonian constraint must be). Thus,
it is clear that the main steps outlined above may be carried out for
{\it any} theory which has the same kinematics, whether or not it has
a Hamiltonian, or Hamiltonian constraint. What must be different
however, is the interpretation to be attached to the calculation if
the theory is not the black hole sector of general relativity (or any
other gravitational theory with black hole solutions for that matter).

This may be illustrated with a concrete example: A generally covariant
theory with the same kinematics as general relativity, but having no
black hole solutions, is obtained from an action identical to the bulk
action (\ref{action}) above, but with SU(2) rather than SL(2,C) as the
gauge group \cite{hk}. The (real) covariant dynamical fields are
$e_a^i$ and $A_a^i$, where $i=1,2,3$ is the SU(2) index. The
Chern-Simons surface term may be added to the action to make the
analogy complete. Thus, consider the action
\be 
S[e,A] = {1\over 8\pi \lambda}\int_M 
\left(e^i\wedge e^j\wedge F^k(A) \epsilon_{ijk}\right) + 
 {1\over 8\pi \lambda} {A_S\over 4\pi}\int_{\p M}  \left( A^i\wedge dA^i 
+ {2\over 3} A^i\wedge A^j \wedge A^k\epsilon_{ijk}\right)
\label{action2}
\ee 
on a manifold $M$ with boundary $\p M$. The constant $A_S$ is now
taken to be the area of the 2-surface $S$ obtained by the intersection
of $\p M$ with a spatial surface $\Sigma$ in $M$, (which is an
embedded surface in $M$ on which the induced metric is not
degenerate). The action(\ref{action2}) has a coupling constant
$\lambda$. Its dimension is fixed by assuming that $e_a^i$ is
dimensionless and $A_a^i$ has dimension $({\rm length})^{-1}$, so
$[\lambda] = {\rm time/mass}$ (giving $S$ dimensions of
action). Note that there is no speed of light $c$ in the theory
because the ``spacetime'' metric $g_{ab}= e_a^ie_b^j\delta_{ij}$ is
degenerate with signature $(0+++)$; $e^i_a$ is a dreibein in four
dimensions. Therefore, although the quantum theory has a fundamental
length $l_F = \sqrt{\hbar\lambda}$, it does not have a fundamental
mass or time.

No component of the boundary $\p M$ is a horizon. Indeed, no spacetime
horizon can even be defined because the four-metric is degenerate.
Nevertheless, the boundary condition on $\p M$ required by functional
differentiability of the action is identical to (\ref{bc1}) above, as
may be directly verified.

We are free to define a boundary with more than one component.
Therefore consider spatial slices $\Sigma$ on which there is an inner
boundary $S$ of area $A_S$, and an outer asymptotic region.
Hamiltonian decomposition of this generally covariant SU(2) gauge
theory reveals, identically to the gravitational case, that there are
volume {\it and} surface degrees of freedom, and that the boundary
symplectic structure is that of Chern-Simons theory.  The phase space
variables may be chosen to have the falloff required of asymptotically
flat spacetimes at spatial infinity (the outer boundary). This
completes the classical analogy with the true gravitational
construction of Ref. \cite{aetal}. (Note that the entire analogy may 
be constructed without reference to any covariant action: one may 
consider phase space variables and constraints on a spatial slice with 
inner and outer boundary, and impose any boundary conditions ``by hand.'')

So far it appears that a difference from the gravitational case, apart
from the identically vanishing Hamiltonian constraint \cite{hk}, is
that there is no Immirzi parameter $\gamma$ ambiguity in the theory
(\ref{action2}). This parameter plays an essential role \cite{aetal}
in fixing to $1/4$ the proportionality constant between entropy and
area. In fact, the Immirzi parameter {\it is} present in the canonical
theory arising from (\ref{action2}). To see this, note that  since 
the spatial dreibein $e^{ai}$ is in general invertible (as for 
general relativity), we can construct the usual connection $\Gamma^a_i$ 
via
\be 
 \p_{[a}e_{b]}^i = \epsilon^i_{\ jk}\Gamma^j_{[a}e_{b]}^k.
\ee
Now define the ``scaled'' canonically conjugate variables 
\be 
^\gamma A_a^i := \Gamma_a^i - \gamma K_a^i;\ \ \ \ \ 
^\gamma E^a_i := {1\over \gamma} E^a_i,
\ee
where $K_a^i \equiv \Gamma_a^i - A_a^i$. This is exactly how the
$\gamma$ ambiguity enters in canonical gravity \cite{imm}.  (Note
however that the $\gamma$ parameter can arise directly by introducing
it into an action via a ``$\gamma$ dual'' of the four-dimensional spin
connection \cite{holst}; in this sense, $\gamma$ is more natural
in general relativity than in the SU(2) theory (\ref{action2}), where
this cannot be done.)

Now, since the entire theory (\ref{action2}) is the kinematical sector
of general relativity, all the above quantization steps are {\it
identical}. Therefore one can associate an entropy with the surface
$S$ which is proportional to its area $A_S$. Furthermore, $\gamma$
may be fixed, as in \cite{aetal}, to get precisely $S=A/4$ in units of 
$l_F$. 

This result seems a bit surprising, since as noted above there is no
speed of light $c$ in (\ref{action2}). However, as noted above, there 
are still two coupling constants $\lambda$ and $\hbar$  in the quantum 
theory which may be combined to give a fundamental length.

The entropy may be interpreted similarly to the so called
``entanglment entropy'' calculated in any system, obtained by tracing
over a portion of the available microscopic degrees of freedom
\cite{entang}. Here as in \cite{aetal}, the bulk states are traced
over to obtain an effective density matrix for the surface states. One
important difference from other entanglement entropy calculations arises 
due to the quantum discreteness arising from the quantization: the 
``polymer'' nature of the discrete geometry \cite{abhpoly} naturally 
makes the entanglement entropy finite.

(II) In the spatial metric ($q_{ab}$) and extrinsic curvature ($K_{ab}$) 
variables for 3+1 gravity, a spatial two-surface $S$ with spatial unit 
normal $s^a$ is marginally outer trapped \cite{hawkell} (ie. is an 
apparent horizon) if 
\be
(q^{ab} - s^as^b)(K_{ab} + D_as_b)=0,
\label{aheqn}
\ee 
where $D_a$ is the covariant derivative associated with $q_{ab}$.  
This equation expresses the statement that the future outward
expansion of light rays vanishes on $S$. In terms of the conjugate
momentum $\tilde{\pi}^{ab} \equiv \sqrt{q}(K^{ab}-K q^{ab})$, the
equation may be succinctly written as
\be 
\tilde{\pi}^{ab}s_as_b = \sqrt{q} D_as^a. 
\ee

This phase space condition has non-trivial solutions on generic
spatial slices of black hole spacetimes. At late times, the radius of
the apparent horizon determined by this equation coincides with the
radius of the black hole event horizon. Therefore, it is the natural
phase space condition to impose on an inner spatial boundary in the
Hamiltonian theory, (if one wishes to follow the route of identifying
black hole entropy as an entropy associated with horizon surface 
degrees of freedom, along the lines of \cite{carlip} or \cite{aetal}).

We now ask whether the condition (\ref{bc1}) is the same as the
apparent horizon equation (\ref{aheqn}).  A straigtforward argument
shows that the answer is no.  The main point to note is that the
apparent horizon equation (\ref{aheqn}) depends on the extrinsic
curvature of $S$ as embedded in a spatial surface $\Sigma$. On the
other hand, eqn. (\ref{bc1}) contains information only about the
extrinsic curvature of $\Sigma$ through $A_a^i$; no derivatives of the
spatial normal $s^a$ of $S$ appear. Indeed, information about $s^a$ in
(\ref{bc1}) only appears in the projection $(q^{ab} - s^as^b)$, and in
the area 2-form $\underline{\sigma}_{ab}$. Now, because no information
about the extrinsic curvature of $S$ as embedded in $\Sigma$ is
present in (\ref{bc1}), this phase space condition does not contain
information about the expansion of light rays on the
surface. Therefore it cannot be the apparent horizon equation. Thus,
the fact that $S$ is taken to be a trapped surface in \cite{aetal}
does not actually enter the quantum black hole entropy calculation.

(III) There is at least one other boundary condition, induced by
adding a different surface term to the gravitational action, which
also leads by arguments similar to the above, to an entropy
proportional to area. Consider the action
\be 
S[e,A] = {1\over 8\pi G_N}
\int_M {\rm Tr}\left[ e\wedge e\wedge F(A)\right] 
+ {1\over 8\pi G_N}{A_S\over 4\pi} 
\int_{\p M} {\rm Tr} \left[ e\wedge F(A)\right].
\label{action3}
\ee 
This is the gravitational action (\ref{action}), but with a ``BF''
theory surface term. The symplectic structure arising from this action
also has a surface contribution, but it is now that of BF theory
rather than Chern-Simons theory. Functional differentialbility 
of the action now requires the condition 
\be 
(\underline{D\wedge e})_{ab} = - {2\pi\over A_S}\ \underline{\sigma}_{ab},
\label{bc2}
\ee   
which fixes the connection on $S$ in a different way than in 
(\ref{bc1}).

The quantization procedure is unchanged, except that the new boundary
condition (\ref{bc2}) must be imposed as a quantum condition. This may
be done in a form identical to (\ref{qbc1}), with the
$\underline{\hat{\sigma}}$ part acting on volume states and the Gauss
law part $\underline{D\wedge \hat{e}}$ acting on surface states: 
\be
\left( 1{\otimes} {2\pi\gamma\over A_S} (\underline{D\wedge
\hat{e}})_{ab}\cdot r + \hat{\underline{\sigma}}_{ab} \cdot r
{\otimes} 1 \right)\, \Psi_V {\otimes} \Psi_S\, = 0
\label{qbc2}
\ee
The difference from the Chern-Simons case is that the surface Gauss
law is now generated by $\underline{D\wedge e}_{ab}$ rather than
$\underline{F}_{ab}$. What is unchanged is that the Gauss law has
sources where edges of the volume spin network state puncture the
surface.

The main question now is whether the entropy calculated by counting
spin network states solving this new constraint, subject to their 
being eigenstates of area with eigenvalue within $\pm l_P^2$ of $A_S$, 
also gives an entropy proportional to the surface's area. Let us first 
recall the two ingredients \cite{aetal} in this 
calculation for the Chern-Simons boundary condition: 
(a) for a set of punctures $P=\{j_{p1},\cdots,j_{pn}\}$, the number of 
solutions $N_P$ of (\ref{qbc1}), for a large number of punctures, is 
\be 
   N_P \sim \prod_{j_p \epsilon P} (2j_p + 1),
\label{csdeg}
\ee
and (b) the eigenvalue of the area operator for the set $P$ is 
\be 
  A_P \sim l_P^2\sum_{p\epsilon P} \sqrt{j_p(j_p+1)}.
\ee
The entropy is obtained by counting all sets $P$ compatible with the 
area $A_P$. The ratio $S/A$ is maximized for all $j_p$ equal and of 
spin $1/2$. In this case $S/A = c\ {\rm ln}2/\sqrt{3}$, where $c$ is a  
constant.  

For the boundary condition (\ref{bc2}) to give the same result, the
degeneracy (\ref{csdeg}) must be the same, since the area spectrum is
obviously unchanged. We argue that this is the case: for a single edge
of spin $j$ piercing $S$, the associated degeneracy is at least the
usual $(2j+1)$ from angular momentum considerations.  Now, looking at
the quantum constraints (\ref{qbc1}) and (\ref{qbc2}), it is apparent
that this degeneracy has the same source, namely the action of
$\underline{\hat{\sigma}}_{ab}$ on volume states. Thus the degeneracy
originates on the right hand side of the respective Chern-Simons or BF
Gauss laws. This suggests that it {\it is} the same in the two
conditions, because its source is the same.  A proof of this
conjecture, similar to that for the Chern-Simons case \cite{km}, may
be possible.

Our observations suggest that the interesting framework for
calculating black hole entropy developed in Ref. \cite{aetal} is
general enough to encompass actions and boundary conditions other than
the specific ones considered there. Furthermore, since the condition
that the two-boundary $S$ is a marginally trapped does not enter the
entropy calculation, the result can also hold for any boundary in
$\Sigma$, even one that has trapped regions outside it.

It is possible to apply the general setup, with boundary conditions
arising from an action, to other theories described by connections,
whether or not they are generally covariant, or have local 
dynamics. Unlike the example in (II) above, a purely topological 
example is 4-dimensional BF theory with a boundary Chern-Simons term 
\be 
S =
{1\over 8\pi \lambda}\int_M {\rm Tr}[B\wedge F(A)] +
{1\over 8\pi \lambda} {A_S\over 4\pi} \int_{\p M}  {\rm Tr}\left[
A\wedge dA + {2\over 3} A\wedge A \wedge A\right].
\label{action4}  
\ee 
Here $B$ is a dimensionless SU(2) valued two-form field, and the the
coupling constant $\lambda$ has the same dimension as in
(\ref{action2}).  The same boundary condition (\ref{bc1}) is induced
by requiring a well defined variational principle. However the
kinematics and dynamics are different. The bulk spatial diffeomorphism
constraint is replaced by the flat connection condition, which renders
the bulk part of the theory entirely topological \cite{horobf,vbf}. 
Spin network states are again solutions to the Gauss law, but now the 
flat connection constraint ``collapses'' the spin networks everywhere 
except on non-trivial bulk topology, and on the boundary punctures where
(\ref{bc1}) holds. The Hilbert space divides up into surface and bulk
states as before. The entropy associated with the surface states can
again be calculated to give similar answers.

This outline for BF theory with a Chern-Simons boundary term lends
further support to the statement that one can always ascribe an
entropy to surfaces by counting the number of surface states. It would
be interesting to produce surface boundary conditions, such that when
quantized as in (\ref{qbc1}), the entropy turns out {\it not} to be
proportional to area, whether or not the theory is generally
covariant. (A large class of boundary conditions for the generally
covariant case for theories with connection variables are considered
in ref. \cite{vs}.)

We close with some further comments: (i) Concerning quantization of
area, if the action (\ref{action}) is indeed the relevant one for the
black hole sector of general relativity, then from a path integral
point of view, quantization of the area $A_S$ also follows from the
standard argument for quantization of the Chern-Simons coupling, due
to the transformation properties of the Chern-Simons term under large
(Yang-Mills) gauge transformations. (ii) It would be interesting to
study the general relativity action with the surface term that induces
the apparent horizon equation (\ref{aheqn}). Since this equation
involves the connection rather than the curvature, the boundary term
would not at first sight be gauge invariant. It may be possible to
avoid this if eqn. (\ref{aheqn}) can be rewritten directly in terms of
the curvature. (iii) It appears that kinematical considerations alone
are not sufficient in themselves to justify the association of black
hole entropy with surface entropy, since as we have seen, this may be
done for generally covariant theories with no black hole solutions.
(iv) An argument using the Immirzi parameter $\gamma$ is essential for
obtaining the correct proportionality constant $1/4$ in $S\sim A_S$
\cite{aetal}. In the non-gravitational examples considered here, this 
parameter is also present in  the two theories defined by (\ref{action2}) 
and (\ref{action4}), provided $E^a_i$ is not degenerate. 
If so,  the entropy associated with bounding surfaces can be fixed 
in the same way. 

\bigskip
{\it Acknowledgements} This work was supported  by the Natural Science 
and Engineering Research Council of Canada.

\end{document}